\documentclass[aps,pra,epsfigure,twocolumn,showpacs]{revtex4}
\usepackage{dcolumn}    
\usepackage{bm} 
\usepackage{graphicx}
\usepackage{amsmath}    
\usepackage{latexsym}
\usepackage{amsfonts}   
\usepackage{amssymb}
\usepackage{array}      
\usepackage{epsfig}
\usepackage{txfonts}
\usepackage{color}
\usepackage{subfigure}
\usepackage{epstopdf}
 
\newcommand{\ket}[1]{\left\vert#1\right\rangle}

\newcommand{\bra}[1]{\left\langle#1\right\vert}

\begin{document}
\title{Quenching small quantum gases: Genesis of the orthogonality catastrophe}
\author{Steve Campbell,$^{1,2}$ Miguel \'Angel Garc\'ia-March,$^{1,3}$ Thom\'as  Fogarty,$^{1,4}$  and Thomas Busch~$^1$}
\affiliation{$^1$Quantum Systems Unit, Okinawa Institute of Science and Technology Graduate University, Okinawa, 904-0495, Japan\\
$^2$Centre for Theoretical Atomic, Molecular, and Optical Physics, School of Mathematics and Physics,
Queen's University Belfast, BT7 1NN, United Kingdom\\
$^3$Departament d'Estructura i Constituents de la Mat\`eria, Universitat de Barcelona, Barcelona, Spain\\
$^4$Theoretische Physik, Universit\"at des Saarlandes, D-66123 Saarbr\"ucken, Germany
}

\begin{abstract}
We study the dynamics of two strongly interacting bosons with an additional impurity atom trapped in a harmonic potential. Using exact numerical diagonalization we are able to fully explore the dynamical evolution when the interaction between the two distinct species is suddenly switched on (quenched). We examine the behavior of the densities, the entanglement, the Loschmidt echo and the spectral function for a large range of inter-species interactions and find that even in such small systems evidence of Anderson's orthogonality catastrophe can be witnessed.
\end{abstract}
\date{\today}
\maketitle

\section{Introduction}
\label{Intro}
Ultracold quantum gases are currently a leading candidate for studying interesting quantum phenomena in interacting many-body systems. One of their key features is the co-existence of a large set of high level experimental control techniques with unmatched isolation from unwanted environmental noises. For this reason they have already been used for experimental implementations of quantum information protocols~\cite{qinformation}, as well as quantum simulators of many condensed matter systems~\cite{qsimulators}. In addition, the physics of these systems is interesting in its own right and small ensembles of ultracold atoms have been used to study non-equilibrium dynamics~\cite{cradle}, interferometry~\cite{Fogarty:2013}, and other fundamental phenomena~\cite{fazio,Goold:2011,ossipov,Deuretzbacher}. 

Similarly the study of multicomponent and hybrid systems is currently raising a great deal of interest. Although inherently more complicated, mixtures of ultracold quantum gases significantly extend the range of new quantum phenomena and  quantum states of matter that can be studied. One recent example are studies in non-Markovianity, where a range of environmental effects can be simulated by controlling the scattering properties between the two components of a quantum gas~\cite{Haikka:2011}. 

The establishment of correlations between the multi-components can also lead to interesting effects, and one of the most striking is the existence of phase separation processes beyond the ones known from mean-field Gross-Pitaevskii physics~\cite{chien,GarciaMarch:2013,GarciaMarch:2013b}. For example, it was recently shown that  in bosonic mixtures in  one-dimensional (1D) traps the presence of a large and repulsive inter-component interaction leads to a composite fermionized state, in which the densities of both 
components overlap while strong anti-correlations between them exist~\cite{Zollner:08a,Hao:09,Hao:09b}. Moreover, increasing then the intra-component interaction in one of the components leads to a sharp crossover to a fully phase separated state~\cite{GarciaMarch:2013b}. 

Other examples of studies on multicomponent systems include the tunneling of a lighter component through the material barrier formed by a second, heavier component~\cite{Pflanzer:2009,Pflanzer:2010}, the miscible-inmiscible dynamics in systems with large particle number imbalances~\cite{sartori} and the dynamics after an interaction quench in an optical lattice~\cite{schmelcher}. Furthermore, the analytical study of the static properties of small ensembles of atoms, including mixtures with up to three atoms has recently received large attention~\cite{Busch:98,Idziaszek:06,Kestner:07,Liu:10,Blume:12,Gharashi:12,Harshman:12,DAmico:13,Harshman:13,Sowinski:13,Volosniev:13,Wilson:13}. Here, we go beyond the aforementioned works and look at the exact dynamics after a sudden switch in the interaction between two distinct species. This will allow us to understand the  dynamical features and correlations in such a system, and give insight into the emergence of the fundamental many-body phenomena of Anderson's 
orthogonality catastrophe (OC). To this end we will quantify the many-body excitations of the system through entanglement, via the von Neumann entropy, and the Loschmidt echo, which have been successfully used to probe dynamical instabilities \cite{LE1,LE2}, critical spin systems~\cite{entspec}, and also explore the frequency spectrum through the spectral function of the system \cite{mahan}. The numerical complexity of the exact approach, however, requires us to restrict the system to one-dimensional dynamics. Note that related studies for fermions have also recently been carried out~\cite{plastina,yulia}.

Our presentation is organized as follows. In Sec.~\ref{tools} we motivate and introduce the model we study and in Sec.~\ref{results} we explore the dynamics resulting from a quench in the interaction between the two different components. These include the behavior of the densities of each species, the entanglement between them, the Loschmidt echo and the spectral function. We then show that the obtained results relate to the emergence of Anderson's OC and conclude in Sec.~\ref{conclusion}.

\begin{figure*}[t]
\begin{center}
\begin{tabular}{ccc}\hspace{-0.2cm}
\includegraphics[scale=0.42]{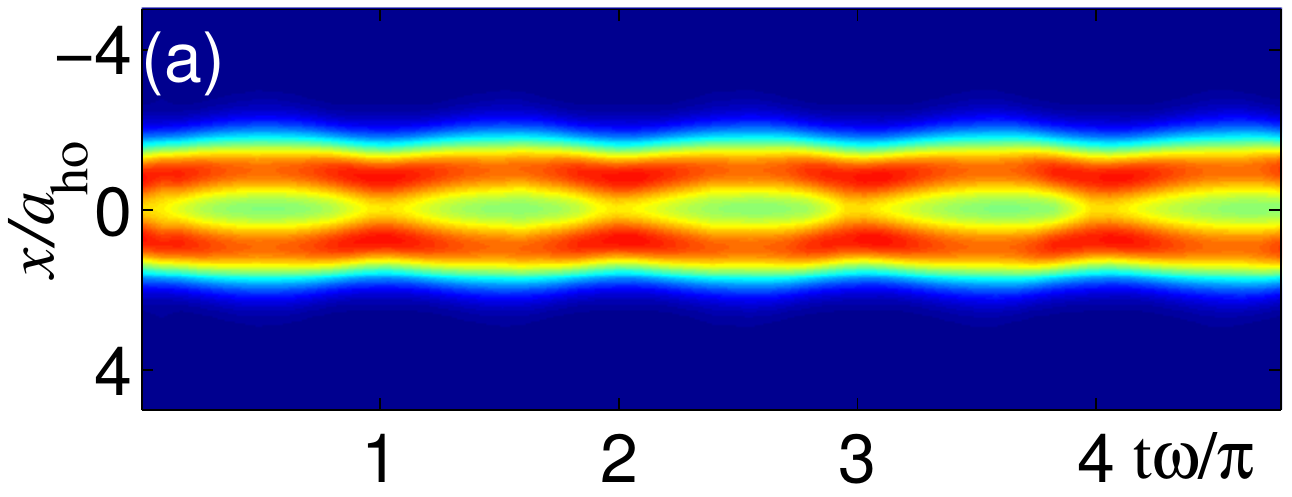} \hspace{-0.0cm}
\includegraphics[scale=0.42]{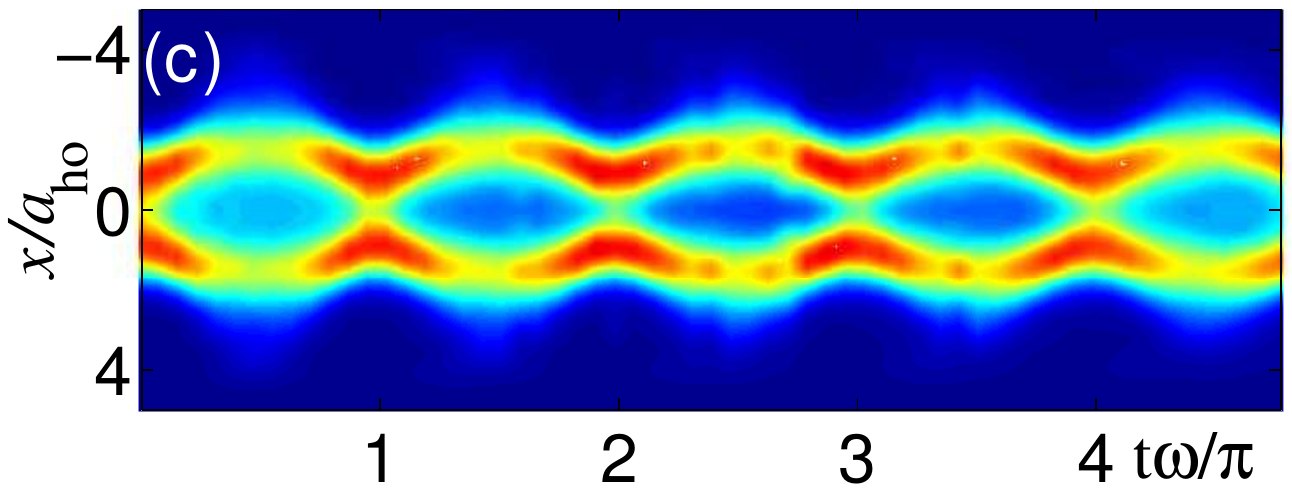} \hspace{-0.0cm}
\includegraphics[scale=0.42]{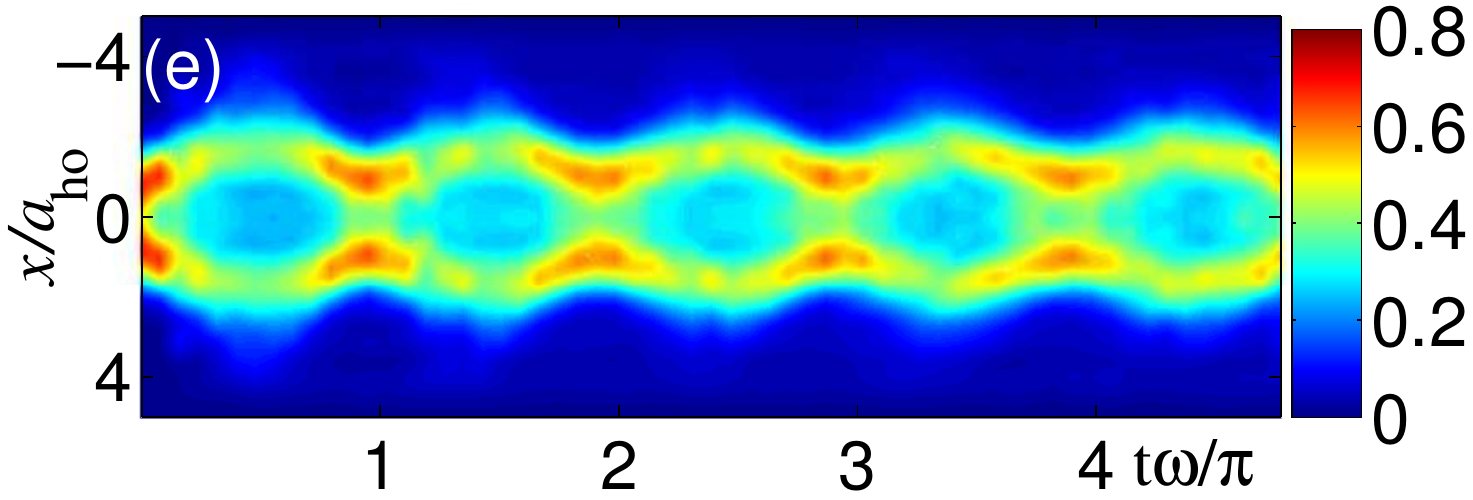}\\ \hspace{-0.1cm}
\includegraphics[scale=0.42]{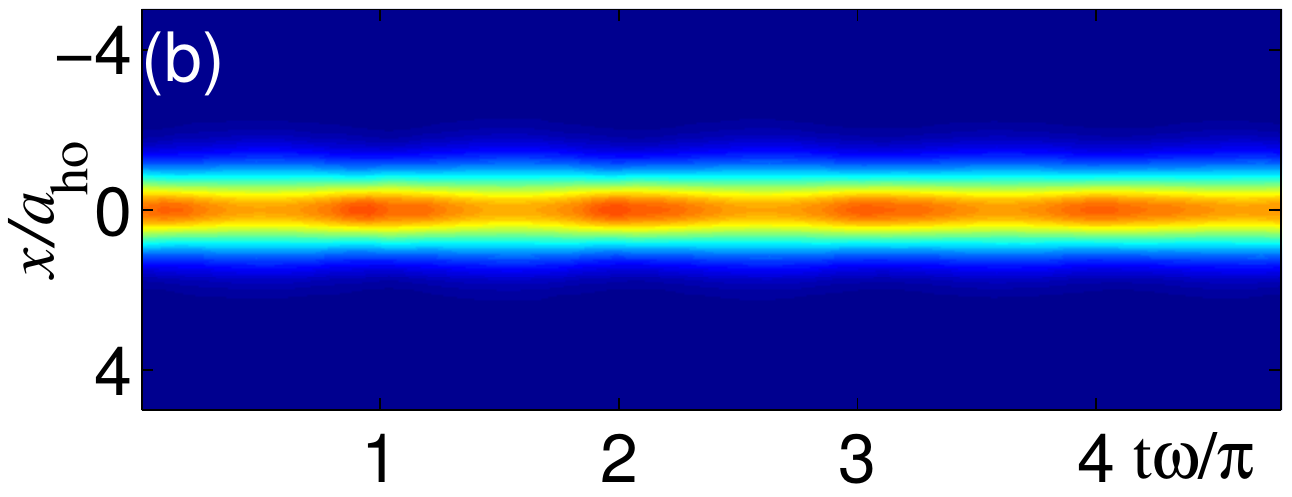} \hspace{-0.0cm}
\includegraphics[scale=0.42]{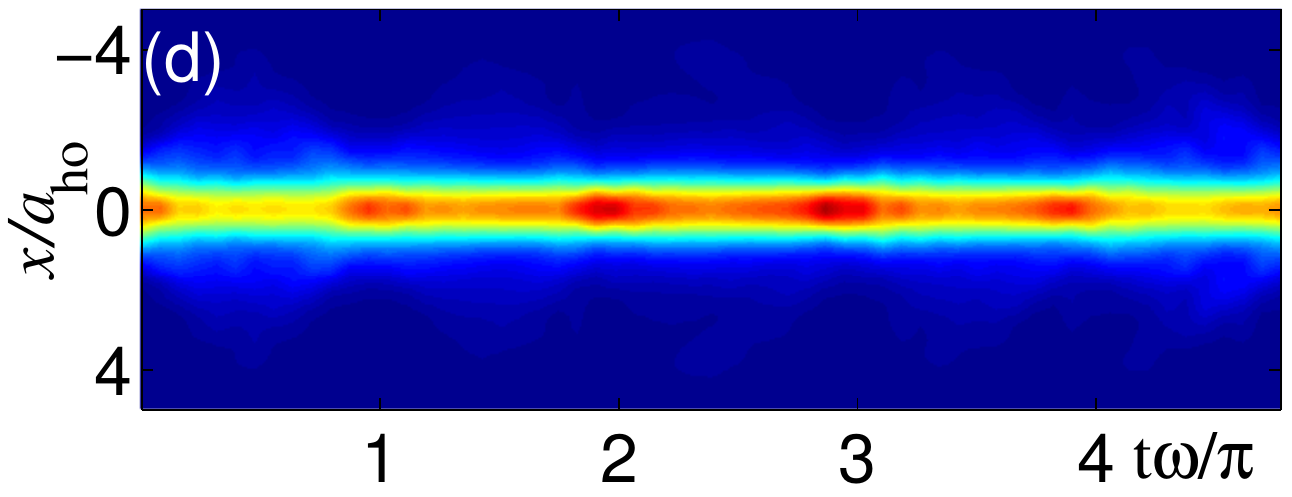} \hspace{-0.0cm}
\includegraphics[scale=0.42]{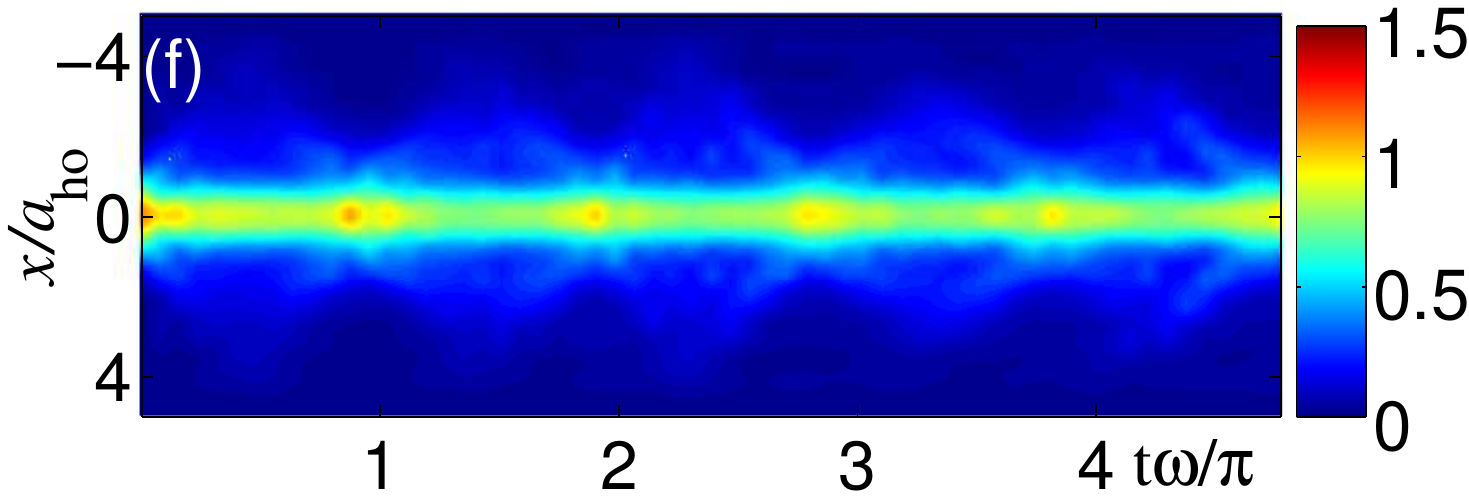} \hspace{-0.0cm}
\end{tabular}
\caption{(Color online) Evolution of the densities after a quench in $g_\mathrm{AB}$. (a) and (b) show the densities for component A and B, respectively, for $g_\mathrm{AB}=0.05g_\mathrm{A}$.  (c) and (d) show the densities for  $g_\mathrm{AB}=0.25g_\mathrm{A}$.  (e) and (f) show the densities for $g_\mathrm{AB}=0.5g_\mathrm{A}$. In all cases,  $g_\mathrm{A}=25a_\mathrm{ho}\hbar\omega$.}
\label{fig1}
\end{center}
\end{figure*}

\section{The model}
\label{tools}
We consider a 1D harmonic trap, which contains a small number of  ultracold, bosonic atoms of species A ($N_{\mathrm{A}}$) and a single atom of a second component, B. Such a mixture can be described by the   many body Hamiltonian  $\mathcal H= \mathcal H_\mathrm{A}+\mathcal H_\mathrm{B} +\mathcal H_\mathrm{AB} $, with
\begin{subequations}
\begin{align}
&  \mathcal H_\mathrm{A}=\sum_{j=1}^{N_\mathrm{A}}\left[-\frac{\hbar^2}{2m}\frac{\partial^2}{\partial x_j^2}+V(x_j) \right]+\sum_{j<j'}^{N_\mathrm{A}}v_{\mathrm{int}}^\mathrm{A}(x_j,x_{j'}),\\
&  \mathcal H_\mathrm{B}=-\frac{\hbar^2}{2m}\frac{\partial^2}{\partial y^2}+V(y),\\
 & \mathcal H_\mathrm{AB} =\sum_{j=1}^{N_\mathrm{A}}  v_{\mathrm{int}}^{AB}(x_j-y), \label{eq:H}
\end{align}
\label{eq:Hamiltonian}
\end{subequations}
where the positions of the atoms in component $\mathrm{A}$ are given by the coordinate $x_j$ and the position of the single atom in component B by $y$.  We assume that both  components have the same mass  $m$, and that they are trapped in the same harmonic potential $V(x)=\frac{1}{2}m\omega^{2}x^{2}$. At low temperatures we can assume contact interactions between the atoms in component A, $v_{\mathrm{int}}^\mathrm{A}=g_\mathrm{A}\delta(x_j-x_{j'}) $, and between species A and the single atom in $\mathrm{B}$, $ v_{\mathrm{int}}^\mathrm{AB}=g_\mathrm{AB}\delta(x_j-y)$. As usual,  $g_\mathrm{A}$  and  $g_\mathrm{AB}$ are the 1D intra- and inter-species coupling constants respectively, and can be tuned independently by means of Feshbach or confinement induced resonances~\cite{Olshanii1998}. In the following, we use harmonic oscillator units and scale all lengths by  $a_\mathrm{ho}=\sqrt{\hbar/(m\omega)}$ and all energies by $\hbar\omega$. Note that if the mass of particle B were much larger or if it was trapped 
in a much tighter trapping potential, its kinetic energy could be neglected and the above model can be reduced to a single component being trapped in a delta-split harmonic oscillator \cite{bus1,bus2,bus3}.
 
To investigate the dynamics given by Hamiltonian \eqref{eq:Hamiltonian} we employ an exact numerical diagonalization algorithm, whose details are given in ~\cite{GarciaMarch:2013}. However, since the numerical resources required grow exponentially with the numbers of particles involved, in the following we will limit ourselves to small systems and consider $N_\mathrm{A}=2$. This drawback is offset by the fact that the exact diagonalization method allows us to investigate quantities beyond the standard mean-field, for example classical and quantum correlations. Also recent experiments have developed unprecedented control to trap small samples of atoms where the exact number of particles can be precisely chosen~\cite{He:10,Serwane:11,Wenz:13,Bourgain:13}. In particular the experimental set-up of Ref.~\cite{will} closely resembles the setting considered here and demonstrates many of the necessary ingredients required to realize the proposed system. With these advances it is now feasible to investigate small ensembles like the one we discuss in this work, which would also allow one to observe the differences between using an odd or an even number of particles in such systems \cite{bus2,Comment}.

\section{Dynamically evolving quantum gas mixtures} 
\label{results}
Since we are especially interested in the dynamics of the correlations in the system, we will in the following focus on the situation where the two atoms of species $\mathrm{A}$ are strongly correlated, i.e.~experience a strong repulsive interaction, $g_{\mathrm{A}}=25a_\mathrm{ho}\hbar\omega$. This is sufficient to ensure they are effectively in the Tonks-Girardeau (TG) limit~\cite{Girardeau:01,Deuretzbacher:07}, while initially completely decoupled from the atom of species B. At $t=0$  a repulsive interaction between the single atom and the diatomic TG gas is switched on and the whole system is allowed to evolve in the harmonic trap. We examine the ensuing non-equilibrium dynamics over a wide range of interspecies interaction strengths, $g_{\mathrm{AB}}\in[0,g_{\mathrm{A}}]$, which cover the transition between the weakly and the strongly correlated regimes. Note that when $g_{AB}=g_A=25\hbar\omega a_{ho}$  both interactions, i.e. between the two particles in A and between the particles of A and the single particle of B, are large enough to guarantee that the system is in the infinite interaction TG limit~\cite{Deuretzbacher:07}.

\subsection{Densities}
We begin by studying how the densities for each species evolve after the inter-species interaction has been switched on. In Figs.~\ref{fig1} (a) and (b) we show the time-evolution of the densities for a weak quench that adjusts the interaction to $g_\mathrm{AB}=\!0.05 g_\mathrm{A}$. We see that the distribution of species $\mathrm{A}$ is initially of TG form and the one for the atom of species $\mathrm{B}$ is Gaussian. While the comparatively small interaction between $\mathrm{A}$ and $\mathrm{B}$ has only a small effect on the density profiles, the repulsion between components A and B still leads to oscillations of the particle pair of species A away and towards the centre, while the single atom of $\mathrm{B}$ remains in the trap centre. The effect on $\mathrm{B}$ is a periodical squeezing in the density, so that it becomes more localized when the 2 atoms of A are closer to each other. The whole process conserves the symmetry of the initial state.

Increasing $g_\mathrm{AB}$ further enhances this behavior and in Figs.~\ref{fig1} (c) and (d) we show the density dynamics for $g_\mathrm{AB}=0.25g_\mathrm{A}$.  Even though the harmonicity of the external potential is now significantly disturbed by the intercomponent interaction, the density evolution can still be seen to be approximately periodic, with a fine-structure appearing for longer times. The two atoms in $\mathrm{A}$ separate further than before due to the increased repulsive interaction with the single atom, which in turn stays mainly localised in the trap centre and loses its Gaussian shape. In fact, it stays strongly localised with small, but extensive wings developing each oscillation period. Comparing this behaviour to the case where the kinetic energy of particle B is neglected (i.e.~where it is modelled by a delta-function potential in the trap centre) shows that the feedback of the A atoms on B plays a significant role in the dynamical evolution, with effects being visible at short time and length scales.

Finally, panels (e) and (f) show the densities when the two species interact strongly with $g_\mathrm{AB}=0.5g_\mathrm{A}$. The qualitative behavior of the interacting atoms remains the same as before and the increased repulsive interaction with the single atom forces the $\mathrm{A}$ atoms to repel even further from the center of the trap. Though qualitatively the dynamics  here look similar for all interactions strengths, we will show below that there are profound differences in the behavior for $g_\mathrm{AB}$ smaller or larger than $\approx0.4g_\mathrm{A}$. 

\begin{figure}[b]
\includegraphics[width=\columnwidth]{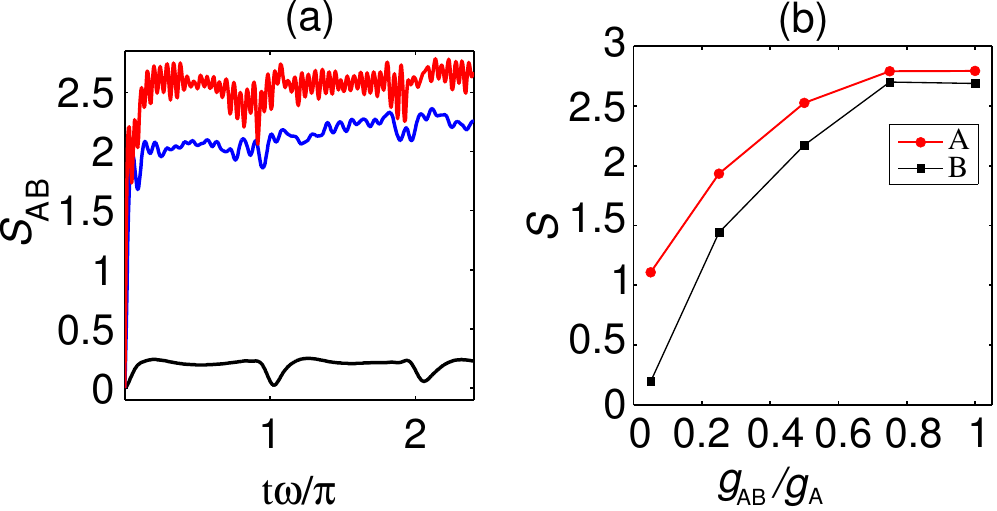}
\caption{(Color online) (a) vNE between A and B as a function of time, when $g_\mathrm{A}=25a_\mathrm{ho}\hbar\omega$ and  $g_\mathrm{AB}=(0.05,0.5,1)g_\text{A}$  for the black (lowest), blue (middle) and red (top) lines, respectively. (b) vNE of the reduced state of species A, red (lighter) line, and species B,  black (darker) line, as a function of $g_\mathrm{AB}/g_\mathrm{A}$.}
\label{fig2}
\end{figure}

\subsection{von Neumann Entropy}
Interactions between the two species will necessarily lead to correlations between them and in the following we will investigate the behaviour of the entanglement created in the system. If the total state is pure, as is the case for our system, the entanglement between two components can be quantified using the von Neumann entropy (vNE) which is found by tracing out one component of the system such that
\begin{equation}
\label{vNEEq}
S(\rho)=-\text{Tr}[\rho~\text{log}_2~\rho] =-\sum_i \lambda_i \text{log}_2 \lambda_i\;.
\end{equation}
Here $\rho$ is the density matrix of the reduced state of one component and $\lambda_i$ are its eigenvalues. For our purposes the vNE between the two components, $S_\mathrm{AB}$, is most conveniently calculated from the single atom of B as its reduced state corresponds exactly to the reduced single particle density matrix (RSPDM). In Fig.~\ref{fig2} (a) we show the behavior of $S_\mathrm{AB}$  for three exemplary values of the inter-species interaction. For small $g_\text{AB}=0.05g_\text{A}$ (black line) we see that a small amount of entanglement is generated, the numerical value of which periodically dips to almost zero and revives with approximately the trap frequency. As $g_\mathrm{AB}$ increases the qualitative features remain, however the amount of entanglement quickly increases to large values and 
performs fast small amplitude oscillations around an almost stationary state value. One can see that for  larger interaction the dips at the approximate trapping frequencies become less prominent and in particular do not reach zero. This can be understood by realizing that the decrease in entanglement comes from an approximate re-focussing on the initial product state of the two component system at multiples of the trapping frequency. This is a feature of the harmonic trapping potential, however with stronger interactions the system becomes increasingly anharmonic and this feature fades away. 

\begin{figure}[t]
\includegraphics[width=\columnwidth]{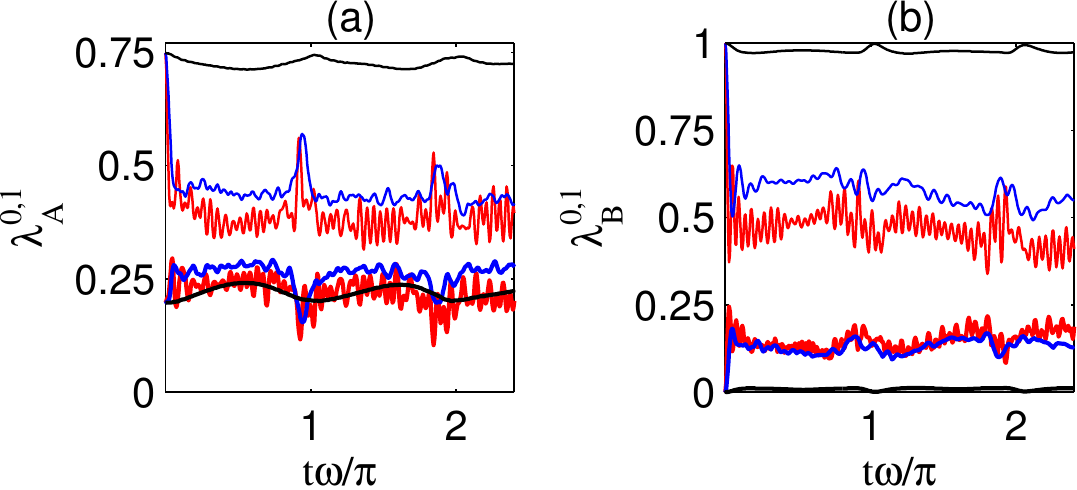}
\caption{(Color online) Natural orbital occupations $\lambda^{0}$ [upper three thin lines] and $\lambda^{1}$ [lower three thick lines]  for (a) species A and (b) species B. Each different color (shade) corresponds to the same color convention as in Fig.~\ref{fig2} (a), i.e. $g_\mathrm{AB}=(0.05,0.5,1)g_\text{A}$ for black, blue and red lines respectively and $g_\mathrm{A}=25a_\mathrm{ho}\hbar\omega$.}  
\label{fig3}
\end{figure}

This behavior is mirrored in the occupation numbers of the natural orbitals of both species, which are given by the eigenvalues of the RSPDMs. In  Fig.~\ref{fig3} we show the largest two values corresponding to the lowest and first excited natural orbital for species A in panel (a) and for species B in panel (b). For small interactions (black lines) the atoms in A are still effectively a TG gas, hence the occupations remain fairly constant, while the B atom continues to  occupy predominantly a single orbital. As the interactions are increased the occupations are affected more significantly: the occupation numbers of the ground and first excited state of the TG pair are moving closer to each other, as the two particle are now experiencing an effective potential that includes the interaction with atom B. This is consistent with the behavior for two atoms in a delta-split trap, where the two lowest lying eigenstates become degenerate if the interaction strength of the barrier goes to infinity \cite{bus1,bus2}. 
At the same time the correlations with the A atoms affect the occupation numbers of the B atom and a finite occupation of the first excited orbital indicates the loss of coherence due to the interaction and also the creation of entanglement. In the limit of $g_\mathrm{AB}\rightarrow g_{\mathrm{A}}$ the two subsystems have features common with a three atom TG gas, as there are very large repulsive interactions between all three atoms. However, the fact that species B is distinguishable from species A, leads to differences from a single component TG gas \cite{Harshman:12,Volosniev:13}, as the symmetrization requirement is different. Nevertheless, in our case the system goes into a highly entangled state in this limit, and tracing out any two particles results in the same amount of entanglement. This can be seen in Fig.~\ref{fig2} (b), where we show the entropies (averaged over time between the trap revivals in the interval $1.25<t\omega/\pi<1.75$) obtained from the RSPDM for a particle of species A (by tracing 
out one particle of A and the single particle of B) and from the RSPDM for particle B (when both particles of species A are traced out) as a function of $g_\mathrm{AB}$. For large values of $g_\text{AB}$ the difference between the entropies approaches zero. 

\begin{figure}[t]
\includegraphics[width=\columnwidth]{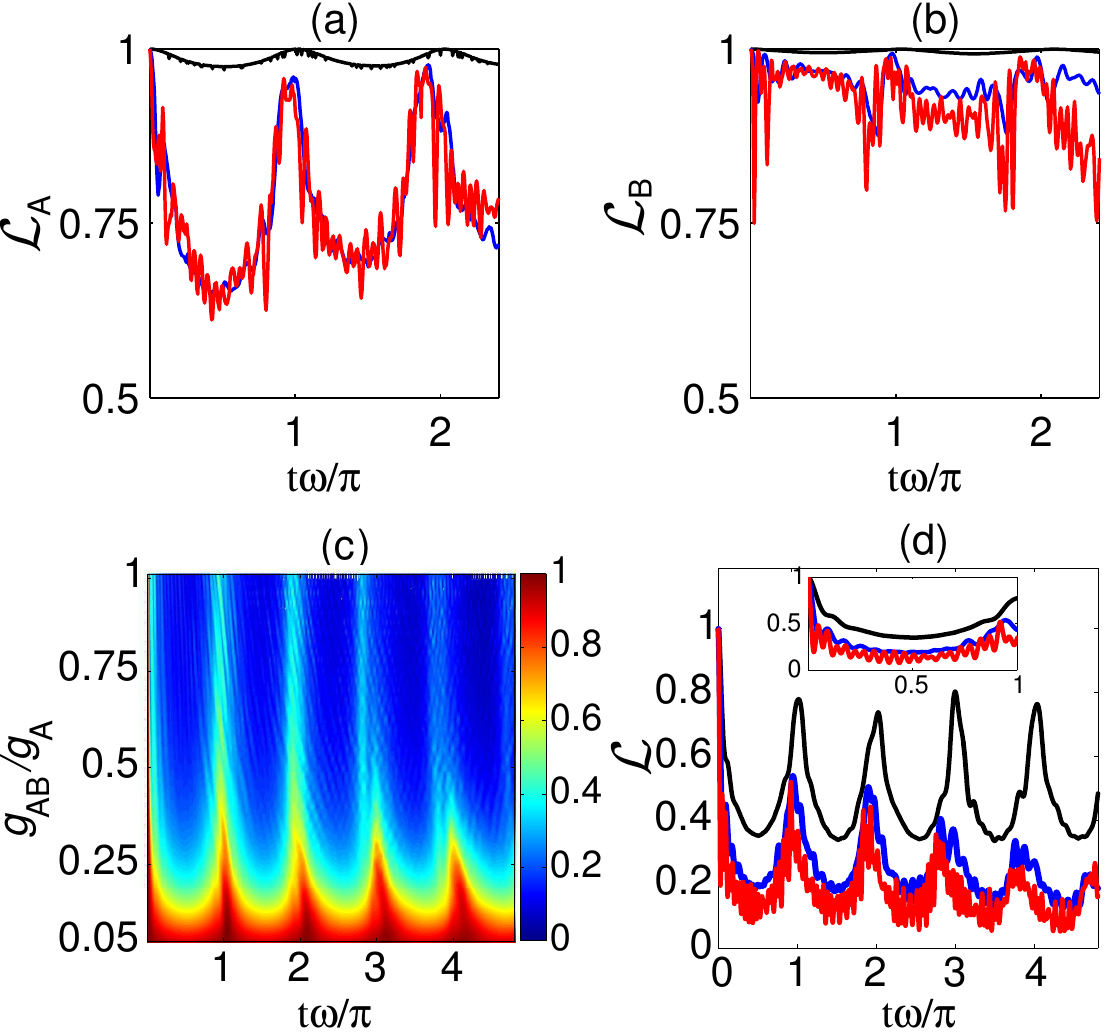}
\caption{(Color online). (a) and (b) Time-evolution of the LE of the individual species for A and B, when $g_\mathrm{AB}=0.05 g_{\mathrm{A}}$ (topmost black line), $0.5 g_{\mathrm{A}}$ (darker blue line), and $g_{\mathrm{A}}$ (lighter red line). (c) Time-evolution of the LE for the total system versus time for the whole range of  $g_\mathrm{AB}\in[0,g_{\mathrm{A}}]$. (d) same as (c) for three particular values $g_\mathrm{AB}=0.25 g_{\mathrm{A}}$ (topmost black line), $0.5 g_{\mathrm{A}}$ (middle blue line) and $g_{\mathrm{A}}$ (bottom red line). Inset shows a zoom over the first period. In all panels $g_\mathrm{A}=25a_{\mathrm{ho}}\hbar\omega$.}
\label{fig4}
\end{figure}

\subsection{Loschmidt Echo}
The Loschmidt echo (LE) is a metric for quantifying the reversibility of a given dynamical evolution. In its original formulation the measure was used to assess the differences in states arising due to imperfect time reversal and it is therefore closely linked with the notion of the thermodynamic arrow of time. Measuring the LE is well within the current experimental capabilities, in particular with NMR setups and recently a closely related quantity, the coherence echo, has been experimentally studied in cold atom systems~\cite{echoexp}. It is also a valuable tool in studying the non-equilibrium thermodynamics of quantum systems, as its definition is closely related to the characteristic function of the work distribution for quenched systems~\cite{thermo}. More recently it has proven extremely useful in understanding decoherence in complex systems~\cite{sindona} and was also used to study phenomena such as quantum phase transitions~\cite{Haikka:2012} and Anderson's OC~\cite{Goold:2011,sindona}. It is defined as
\begin{equation}
\label{EchoEq}
 \mathcal{L}(t)=\vert \bra{\Psi_0} e^{i\mathcal{H}_it}e^{-i\mathcal{H}_ft}\ket{\Psi_0}\vert^2,
\end{equation}
where $\Psi_0$ and $\mathcal{H}_i$ are the initial state and Hamiltonian, respectively, and $\mathcal{H}_f$ describes the Hamiltonian after a quench. In our case, the LE allows us to assess how the dynamics changes when the interaction between the two components is switched on. Therefore the initial Hamiltonian is simply the free evolution of the two components with no interaction between them, i.e. $g_{\mathrm{AB}}=0$, while the final Hamiltonian $\mathcal{H}_f$ is the one which incorporates the finite inter-species interaction.

To study the LE of the individual species, the expression in Eq.~\eqref{EchoEq} cannot be used, as we do not have the wavefunctions of the individual species available. It is possible however to calculate the LE using the reduced states of the composite system as these contain all of the information of the dynamics of $\Psi(t)$. In this way the LE is defined as
\begin{align}
\mathcal{L}_{\mathrm{A,B}}(t)&=\sum_m \omega_m \vert \langle \psi_m \vert e^{i \rho t}e^{-i \rho't}\vert \psi_m \rangle \vert^2\nonumber\\
&=\sum_m \omega_m \Bigg[ \Bigg( \sum_n \cos(\omega'_{n} t) \vert \langle \psi_m \vert \phi'_{n}\rangle \vert^2 \Bigg)^2\nonumber\\
&\qquad\qquad+  \Bigg( \sum_n \sin(\omega'_{n} t) \vert \langle \psi_m \vert \phi'_{n}\rangle \vert^2 \Bigg)^2 \Bigg]\;,
\end{align}
where $\rho$ and $\rho'$ are the reduced density matrices of species A or B from the respective Hamiltonians $\mathcal{H}_i$ and $\mathcal{H}_f$ with corresponding eigenvalues $\omega_m$ and $\omega'_n$ and eigenvectors $\psi_m$ and $\phi'_n$~\cite{LiLi}.

In Figs.~\ref{fig4}(a) and (b) we show the LE of the individual species as a function of time after the quench for three distinct values of the inter-species interaction strength. For small $g_\text{AB}$ the LE does not deviate much from unity for both species, but for larger interaction the dynamics of each species can be seen to be quite different. 
Periodic revivals, which occur with roughly the trap frequency, become visible in $\mathcal{L}_\mathrm{A}$, whereas at the same points the LE deviates from unity for species B. This confirms the earlier observation that the main effect of the quench on the state of B is to get squeezed whenever the two A atoms come close. Note that this behavior shows that the correlations that are built up in the entanglement between the two components lead mainly to a real-space distortion of component B at multiples of the trap frequency.

In Fig.~\ref{fig4} (c) we show the time-evolution of the LE for the total state of both species for the whole range of $g_\mathrm{AB}$. Again, periodic revivals at multiples of the trap frequency are clearly visible and stronger values of $g_\text{AB}$ quickly lead to a deviation of the LE from 1. A particular sharp drop in the LE can be seen in the interval $g_\mathrm{AB}/g_\text{A}\in[0,0.4]$, after which the LE does not reach any values close to unity anymore and has a minimum value of $0.043$  when $g_\mathrm{AB}/g_\text{A}\!\to\!1$. This indicates that the initial and the evolved states exhibit a vanishing overlap, which can be interpreted as a precursor of the OC that can be expected in larger samples.  At these large interactions the LE also acquires high frequency oscillations, which are shown in Fig.~\ref{fig4} (d) for $g_{\mathrm{AB}}=0.5$ and $1$ and which are also present in $\mathcal{L}_\text{A}$ and $\mathcal{L}_{\mathrm{B}}$ [see Figs.~\ref{fig4} (a) and (b)]. From the inset it can be seen 
that the amplitude of these oscillations increases with increasing $g_\text{AB}$, which indicates that this is related to the increased kinetic energy of $\mathrm{B}$ in the effective potential provided by the A atoms.

Interestingly, even for such a small sized system the qualitative features here are similar to large fermionic systems interacting with a single qubit impurity~\cite{Goold:2011}. This implies that even in small bosonic systems the emergence of Anderson's OC can be witnessed.

\subsection{Spectral Function and Orthogonality Catastrophe}
In condensed matter, observing the OC is often achieved by studying the behavior of the spectral function, which offers insight into the fundamental excitations of a system \cite{mahan}. The spectral function is given by
\begin{equation}
\label{spectral}
A(\omega)=2 \mbox{Re} \int_{-\infty}^{\infty}e^{i\omega t}\nu(t)dt
\end{equation}
where $\nu(t)= \bra{\Psi_0} e^{i\mathcal{H}_it}e^{-i\mathcal{H}_ft}\ket{\Psi_0}$ is the time dependent overlap of the initial and the post quench states which is related to the LE through $\mathcal{L}(t)=\vert \nu(t) \vert^2$. In the case of the OC the spectral function exhibits an asymmetric broadening which decays with a power law distribution highlighting the sudden change of the excitations in the system. We calculate Eq.~\eqref{spectral} by taking a suitably large time window so as to capture all the relevant dynamics occurring in the long-time limit.

\begin{figure}[t]
\includegraphics[width=8.3cm]{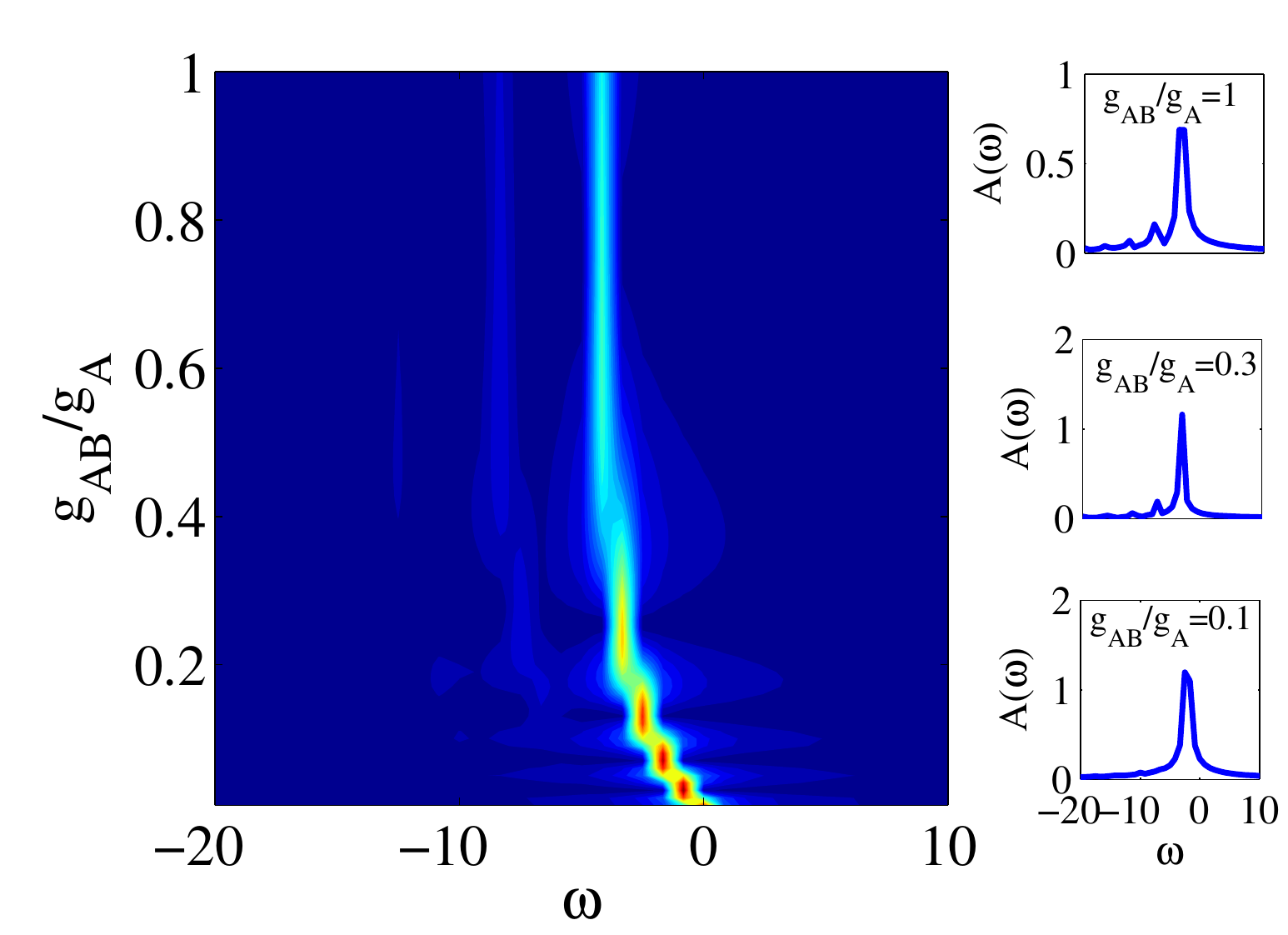}
\caption{(Color online). Spectral function $A(\omega)$ as a function of $g_{\mathrm{AB}}/g_\text{A}$. The quasi-particle peak moves from $\omega=0$ to more negative finite frequencies as $g_{\mathrm{AB}}$ is increased until $g_{\mathrm{AB}}/g_{\mathrm{A}}\gtrapprox0.4$, after which the dynamics of the system does not drastically change. Also visible at more negative finite $\omega$ is a cusp which is a consequence of the oscillating behaviour of species B as seen in Fig.~\ref{fig4}. Cuts of the spectral function are shown for $g_{\mathrm{AB}}/g_{\mathrm{A}}=0.1, 0.3, 1$ and $g_\mathrm{A}=25a_\mathrm{ho}\hbar\omega$.}
\label{fig5}
\end{figure}

In Fig.~\ref{fig5} we show the spectral function for the entire system as a function of the interspecies interaction $g_\text{AB}$. We see that with increasing interaction the quasi-particle frequency shifts away from the origin until it reaches a constant value for $g_\mathrm{AB}\gtrapprox0.4g_\mathrm{A}$ where it then remains. This indicates that the system has gone from a composite two species system to a strongly correlated many-particle system, which has several properties in common with a single component TG gas. Furthermore, we see that the spectrum becomes asymmetric for increased interactions, as the system has been pushed out of equilibrium and tries to settle into a new state. The decay of the spectral function for negative $\omega$ is consistent with that seen in \cite{Goold:2011} and is an indication of the OC. However, in this work the full dynamics of the impurity is taken into account, which can be seen to lead to the appearance of a cusp for 
large $g_{\mathrm{AB}}$. It is the same effect that manifests itself in the high frequency oscillations in the LE of the single particle, $\mathcal{L}_{\mathrm{B}}$ in Fig.~\ref{fig4} (b) and also in the LE of the composite state, $\mathcal{L}$ in the inset of Fig.~\ref{fig4} (d). Similar features have also been observed in Ref.~\cite{Demler} when impurities in fermions approach criticality. 

\section{Conclusions}
\label{conclusion}
We have investigated the dynamics of a small two-component quantum gas, consisting of a strongly interacting Tonks Girardeau pair and a single impurity, after a sudden quench in the interspecies interaction. The resulting repulsion between the impurity atom and the TG pair drives the system out of equilibrium and initiates an interesting and fundamental dynamical evolution. By examining the von Neumann entropy  and the Loschmidt echo we have found distinct low and high frequency dynamics, which stem from the trap and kinetic behavior of the impurity, respectively. Furthermore we have found that the LE acquires near vanishing values for large interspecies scattering, implying the emergence of Anderson's orthogonality catastrophe, which is remarkable as the OC is mainly considered to be a phenomenon only observed in large ensembles. Indeed, in such large ensembles a power-law decay of the spectral function is a clear signature of the OC. While achieving such a power law is not possible due to the small system size considered here, we have shown that a second dominant peak appears in the spectral function visible at large frequencies for suitably large interspecies interaction, and we have proposed that it is due to the dynamic behavior of the impurity particle in the system, which oscillates in an effective potential given by the trap and the TG pair. The resulting asymmetry in the spectral function and the near vanishing value of the LE suggests that even in small bosonic systems the first signs of the OC can manifest themselves. Finally, we expect our results to be applicable to studying in more detail the non-equilibrium thermodynamics of small bosonic systems along the lines of those presented for fermions in~\cite{plastina}.

\acknowledgments 
The authors thank G. De Chiara and M. Paternostro for useful discussions and we acknowledge support from OIST Graduate University. SC is funded through the EU Collaborative Project TherMiQ (Grant Agreement 618074). MAGM acknowledges financial support from the DGI (Spain) Grant No. FIS2011-24154. TF acknowledges funding from BMBF (QuORep, Contract No. 16BQ1011), by the German Research Foundation.


\begin{thebibliography}{99}

\bibitem{qinformation}
M.~Anderlini, P.J.~Lee, B.L.~Brown, J.~Sebby-Strabley, W.D.~Phillips and J.V.~Porto, Nature (London) {\bf 448}, 452 (2007); S.~Trotzky, P.~Cheinet, S.~F\"olling, M.~Feld, U.~Schnorrberger, A.M.~Rey, A.~Polkovnikov, E. A.~Demler, M.D.~Lukin and I~ Bloch, Science {\bf 319}, 295 (2008); O.~Mandel, M.~Greiner, A.~Widera, T.~Rom, T.W.~H\"ansch, and I.~Bloch, Phys.~Rev.~Lett.~{\bf 91}, 010407 (2003).

\bibitem{qsimulators} D. Jacksch, Contemp. Phys. {\bf 45} 367 (2007); I. Bloch and W. Zwerger, Rev. Mod. Phys. {\bf 80} 885 (2008);  M. Lewenstein,  A. Sanpera, V. Ahufinger, B. Damski, A. Sen(De) and U. Sen, Adv. Phys. {\bf 56} 243 (2007); M. Lewenstein,  A. Sanpera, and V. Ahufinger, {\it œUltracold Atoms in Optical Lattices: Simulating quantum many-body systems} Oxford Univ. Press, 2012.


\bibitem{cradle} Toshiya Kinoshita, Trevor Wenger, and David Weiss, Nature {\bf 440} 900-903 (2006); Paolo P. Mazza, Mario Collura, Márton Kormos, and Pasquale Calabrese arXiv:1407.1037.

\bibitem{Fogarty:2013} Thom\'as Fogarty, Anthony Kiely, Steve Campbell, and Thomas Busch, Phys. Rev. A {\bf 87}, 043630 (2013).

\bibitem{fazio} Sebastiano Peotta, Davide Rossini, Marco Polini, Francesco Minardi, and Rosario Fazio, Phys. Rev. Lett. {\bf 110}, 015302 (2013).

\bibitem{Goold:2011} J. Goold, T. Fogarty, N. Lo Gullo, M. Paternostro, and Th. Busch, Phys. Rev. A {\bf 84}, 063632 (2011).

\bibitem{ossipov} A. Ossipov, arXiv:1404.2506

\bibitem{Deuretzbacher} F. Deuretzbacher, D. Becker, J. Bjerlin, S. M. Reimann, and L. Santos, arXiv:1310.3705.

\bibitem{Haikka:2011} P. Haikka, S. McEndoo, G. De Chiara, G. M. Palma, and S. Maniscalco, Phys. Rev. A {\bf 84}, 031602(R) (2011); S. McEndoo, P. Haikka, G. De Chiara, G. M. Palma, and S. Maniscalco, EPL {\bf 101}, 60005 (2013).

\bibitem{chien} Chih-Chun Chien and Fred Cooper, Phys. Rev. A, {\bf 87}, 045602 (2013).


\bibitem{GarciaMarch:2013} Miguel \'Angel Garc\'{\i}a-March and Thomas Busch, Phys. Rev. A {\bf 87}, 063633 (2013).

\bibitem{GarciaMarch:2013b} M. A. Garc\'{\i}a-March, B. Juli\'a-D\'i­az, G. E. Astrakharchik, Th. Busch, J. Boronat, and A. Polls, Phys. Rev. A {\bf 88}, 063604 (2013).


\bibitem{Zollner:08a}
S. Z\"{o}llner, H.-D. Meyer, and P. Schmelcher, 
Phys.~Rev.~A, \textbf{78} 013629 (2008).

\bibitem{Hao:09}
Y.J. Hao and S. Chen, 
Phys.~Rev.~A \textbf{80} 043608 (2009).

\bibitem{Hao:09b} 
Y.J. Hao and S. Chen, Eur. Phys. J. D \textbf{51}, 261-266 (2009).

\bibitem{Pflanzer:2009} A.~C. Pflanzer, S. Z\"{o}llner, and P. Schmelcher, Journal of Physics B: Atomic, Molecular and Optical Physics,   \textbf{42} 231002 (2009).

\bibitem{Pflanzer:2010} A.~C. Pflanzer, S. Z\"{o}llner, and P. Schmelcher Phys.~Rev.~A, \textbf{81} 023612 (2010).

\bibitem{sartori} Alberto Sartori and Alessio Recati, Eur. Phys. J. D, 67:260, (2013).

\bibitem{schmelcher} S.I. Mistakidis, L. Cao, and P. Schmelcher, arXiv:1404.7840

\bibitem{Busch:98} T. Busch, B.-G. Englert, K. Rzazewski, and M. Wilkens, 
Found. Phys.~\textbf{28}, 549  (1998).

\bibitem{Idziaszek:06} Z. Idziaszek and T. Calarco Phys. Rev. A ~\textbf{74}, 022712 (2006).

\bibitem{Kestner:07} J. P. Kestner and L.-M. Duan Phys. Rev. A ~\textbf{76}, 033611 (2007).

\bibitem{Liu:10} X.-J. Liu, H. Hu, and P. D. Drummond, Phys. Rev. A ~\textbf{82}, 023619 (2010).

\bibitem{Blume:12} D.~Blume,  Rep. Prog. Phys. \textbf{75}  046401  (2012). 

\bibitem{Gharashi:12} S. E. Gharashi, K. M. Daily, and D. Blume,  Phys. Rev. A ~\textbf{86}, 042702 (2012).

\bibitem{Harshman:12} N. L. Harshman, Phys. Rev. A ~\textbf{86}, 052122 (2012).

\bibitem{DAmico:13} P. D'Amico and M. Rontani, J. Phys. B {\bf 47} 065303 (2014).

\bibitem{Harshman:13} N.L. Harshman,  arXiv:1312.6107 (2013).

\bibitem{Sowinski:13} T. Sowinski, T. Grass, O. Dutta, and M. Lewenstein Phys. Rev. A ~\textbf{88}, 033607 (2013).

\bibitem{Volosniev:13} A. G. Volosniev, D. V. Fedorov, A. S. Jensen, 
M. Valiente, N. T. Zinner arXiv:1306.4610 (2013).

\bibitem{Wilson:13} 
B. Wilson, A. Foerster, C. C. N. Kuhn, I. Roditi, D. Rubeni, Phys. Lett. A {\bf 378}  1065 (2014).


\bibitem{LE1} F. M. Cucchietti, D. A. R. Dalvit, J. P. Paz, and W. H. Zurek, Phys. Rev. Lett. {\bf 91}, 210403 (2003).
\bibitem{LE2} H. T. Quan, Z. Song, X. F. Liu, P. Zanardi, and C. P. Sun, Phys. Rev. Lett. {\bf 96}, 140604 (2006).
\bibitem{mahan} G. D. Mahan, Many Particle Physics (Springer-Verlag, Berlin/New York, 2000).

\bibitem{entspec} G. Torlai, L. Tagliacozzo, and G. De Chiara, J. Stat. Mech. (2014) P06001; Elena Canovi, Elisa Ercolessi, Piero Naldesi, Luca Taddia, and Davide Vodola, Phys. Rev. B {\bf 89}, 104303 (2014).

\bibitem{plastina}A Sindona, J Goold, N Lo Gullo, and F Plastina, New J. Phys., {\bf 16}, 045013 (2014).
\bibitem{yulia} Yulia E. Shchadilova, Pedro Ribeiro, and Masudul Haque, Phys. Rev. Lett. {\bf 112}, 070601 (2014); Yulia E. Shchadilova, Pedro Ribeiro, and Masudul Haque, Phys. Rev. B {\bf 89}, 104102 (2014).

\bibitem{Olshanii1998}
M.~Olshanii, Phys. Rev. Lett. \textbf{81} 938 (1998); 
E. Haller, M. J. Mark, R. Hart, J. G. Danzl, L. Reichs\"ollner, V. Melezhik, P. Schmelcher, H.-C. N\"agerl, Phys. Rev. Lett.  {\bf 104}, 153203 (2010).

\bibitem{bus1}
D. S. Murphy, J. F. McCann, J. Goold, and Th. Busch, Phys. Rev. A {\bf 76}, 053616 (2007).

\bibitem{bus2}
J. Goold and Th. Busch, Phys. Rev. A {\bf 77}, 063601 (2008).

\bibitem{bus3}
J. Goold, M. Krych, Z. Idziaszek, T. Fogarty and Th. Busch, New J. Phys. {\bf 12} 093041 (2010).

\bibitem{He:10} X. He, P. Xu, J. Wang, and M. Zhan, Opt. Express, \textbf{18} 13586 (2010).

\bibitem{Serwane:11}
F.~Serwane, G.~Z\"{u}rn, T.~Lompe, T.B.~Ottenstein, A.N. Wenz, and S.~Jochim, 
Science, \textbf{332} 336 (2011).


\bibitem{Wenz:13} 
A. N. Wenz, G. Z\"{u}rn, S. Murmann, I. Brouzos, T. Lompe, S. Jochim, Science~\textbf{342}, 457 (2013). 

\bibitem{Bourgain:13}  R.~Bourgain, J.~Pellegrino, A. Fuhrmanek, Y.R.P. Sortais, and A. Browaeys,  
Phys. Rev. A~\textbf{88}, 023428 (2013).

\bibitem{will} Sebastian Will, Thorsten Best, Simon Braun, Ulrich Schneider, and Immanuel Bloch, Phys. Rev. Lett., {\bf 106}, 115305 (2011).


\bibitem{Comment}
Going from $N_\mathrm{A}=2$ to $N_\mathrm{A}=3$ is numerically not trivial and the focus of a future work.

\bibitem{Girardeau:01} 
M.D.~Girardeau, E.M. Wright, and J.M. Triscari, 
Phys.~Rev.~A \textbf{63}, 033601 (2001).

\bibitem{Deuretzbacher:07} 
F.~Deuretzbacher, K. Bongs, K. Sengstock, and D.  Pfannkuche, 
Phys.~Rev.~A \textbf{75}, 013614 (2007).

\bibitem{echoexp} M. F. Andersen, A. Kaplan, and N. Davidson, Phys. Rev. Lett. 90, 023001 (2003); M. F. Andersen, A. Kaplan, T. Gr\"unzweig, and N. Davidson, {\it ibid.} 97, 104102 (2006).

\bibitem{thermo} Michele Campisi, Peter H\"anggi, and Peter Talkner, Rev. Mod. Phys., {\bf 83}, 771-791 (2011); Peter Talkner, Eric Lutz, and Peter H\"anggi, Phys. Rev. E {\bf 75}, 050102(R) (2007); R. Dorner, S. R. Clark, L. Heaney, R. Fazio, J. Goold, and V. Vedral, Phys. Rev. Lett. {\bf 110}, 230601 (2013); Laura Mazzola, Gabriele De Chiara, and Mauro Paternostro Phys. Rev. Lett. {\bf 110}, 230602 (2013).

\bibitem{sindona}A. Sindona, J. Goold, N. Lo Gullo, S. Lorenzo, and F. Plastina, Phys. Rev. Lett. {\bf 111}, 165303 (2013).

\bibitem{Haikka:2012} P. Haikka, J. Goold, S. McEndoo, F. Plastina, and S. Maniscalco, Phys. Rev. A {\bf 85}, 060101(R) (2012).

\bibitem{LiLi} Y. C. Li and S. S. Li, Phys. Rev. A {\bf 76} 032117 (2007).

\bibitem{Demler}
M. Knap,  A. Shashi, Y. Nishida, A. Imambekov, D.~A. Abanin, and E. Demler,  Phys. Rev. X. {\bf 2}, 041020 (2012).

\end{thebibliography}
\end{document}